# Asymmetric Nucleation Processes in Spontaneous Mode Switch of Active Matter

Bing Yang (杨冰)[1,2], and Yanting Wang (王延颋)[1,2,*]

[1]CAS Key Laboratory of Theoretical Physics, Institute of Theoretical Physics, Chinese Academy of Sciences, Beijing 100190, China

[2]School of Physical Sciences, University of Chinese Academy of Sciences, Beijing 100049, China

**ABSTRACT:** Flocking and vortical are two typical motion modes in active matter. Although it is known that the two modes can spontaneously switch between each other in a finite-size system, the switching dynamics remain elusive. In this work, by computer simulation of a two-dimensional Vicsek-like system with 1000 particles, we find from the perspective of classical nucleation theory that the forward and backward switching dynamics are asymmetric: from flocking to vortical is a one-step nucleation process, while the opposite is a two-step nucleation process with the system staying in a metastable state before reaching the final flocking state.

Flocking movement (all particles moving roughly parallelly, Fig. 1(a)) and vortical movement (all particles moving rotationally around the center of mass, Fig. 1(b)) are two important motion modes of active matter. The flocking movement is widely found in the migratory process of biological flocks, such as locusts and zebras [1]. Thousands of starlings may defend predators by moving in the flocking mode [2]. A swarm of unmanned aerial vehicles (UAV) can search for victims and track targets with the flocking motion [3,4]. By contrast, the vortical movement is very often a consequence of environmental disturbance or required by biological functions. A colony of bacillus subtilis grows to be in the vortical morphotype as a response to a hostile environment, such as a diffusion-limited growth condition or an infertile agar substrate [5]. Ordemann et al. [6] have studied how a light beam induces Daphnia to move in the vortical mode. Blair et al. [7] have confirmed experimentally that vertically vibrated granular rods can move vortically in a circular container with appropriate packing fraction, vibration amplitude, and frequency. Self-propelled microsensors powered by alternating electric fields can form the vortical state to achieve microfluidic mixture in microchannels [8]. A UAV swarm can implement detecting, monitoring, and loitering functions in the vortical motion mode [4,9].

For a finite-size system, the two motion modes may spontaneously switch between each other. The switch between the flocking and vortical swimming modes of golden shiners in a shallow tank was observed to be influenced by boundary and internal factors [10]. The bistable switching behavior was explained by considering the competition between attractive interaction and alignment [11], or between attractive and repulsive interactions [12]. The influences of alignment, attraction, and blind angle on the rate of mode switching were also investigated [13,14]. Chen and Hou [15] found that a vortex reversed its chirality through a hierarchical process: particles flipped their moving directions layer by layer from the outside to the inside. For a three-dimensional model of active Brownian particles, it has been found that the switching and hysteresis phenomena cannot exist simultaneously [16].

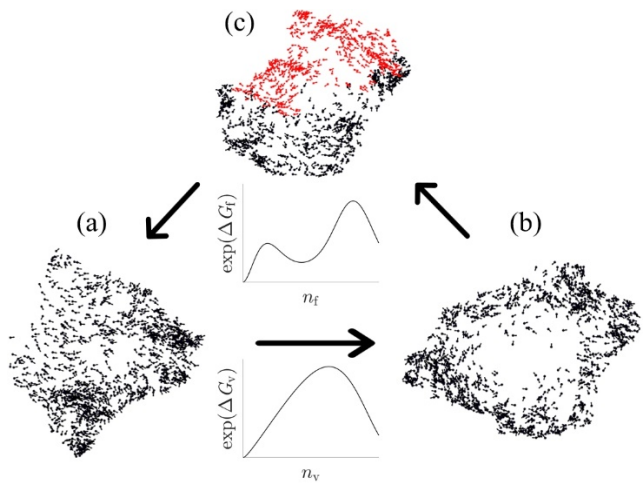

FIG. 1. Spontaneous switch of the system from the flocking mode (a) to the vortical mode (b) is a one-step nucleation process, while the reverse is a two-step nucleation process with a metastable state (c). The particles colored in red in (c) form a flocking nucleus in the vortical background.

The switching dynamics between the flocking and vortical motion modes are essential for understanding the physical nature of active matter, which remains elusive. In this work, we modify the Vicsek model to simulate a flock of active matter with 1000 particles under the free boundary condition. With a certain combination of the orientational force and associated noise, we observe that the system spontaneously switches between the flocking mode and the vortical mode. By employing the classical nucleation theory (CNT) [17-19], we identify that the *forward mode-switching* from flocking motion to vortical motion is a one-step nucleation process with a critical nucleation size of 651 particles, while the *backward mode-switching* from vortical to flocking is a two-step nucleation process when the system goes over the first free energy barrier with a flocking nucleus of 187 particles to reach a metastable state with a flocking nucleus of 440 particles, before it goes over the second barrier with a flocking nucleus of 766 particles to be in the final flocking state.

The two-dimensional active matter is modelled by a Vicsek-like model modified from the Vicsek model [20]. For each particle, an orientational force $F^o(t)$ is applied to mimicking the attempt of adjusting its moving direction parallel to its nearest neighbors [21,22]. A short-range repulsive force $\bar{F}^r(t)$ is adopted to retain the finite exclusive volume of a particle [23], and the over-damped Langevin equation without translational noise [24] is used to eliminate the influence of $\bar{F}^r(t)$ on the self-propulsion velocities. With the free boundary condition applied, an attractive force $F^a(t)$ pointing inwards is applied to the particles on the boundary to prevent them from easily fleeing away [25]. Each particle has a fixed self-propulsion speed of $v_0 = 0.5$, while the direction of the self-propulsion velocity $\theta_i(t)$ at time $t$ is adjusted by $F^o(t)$, $F^a(t)$ if the particle is on the boundary, and a Gaussian white noise $\xi_i(t)$ mimicking the judging error to moving directions:

$$\theta_i(t) = \theta_i(t-\Delta t) + \dot{\theta}_i(t)\Delta t$$
$$\dot{\theta}_i(t) = \alpha F_i^o(t) + \varepsilon F_i^a(t) + \xi_i(t)$$
$$F_i^o(t) = \sum_{j\in \text{VN}(i)} \sin(\theta_j(t-\Delta t) - \theta_i(t-\Delta t))$$
$$F_i^a(t) = (1-u_i^2)\sin(\theta(\bar{r}_i^c(t-\Delta t) - \bar{r}_i(t-\Delta t)) - \theta_i(t-\Delta t))$$
$$\langle \xi_i(t) \rangle = 0$$
$$\langle \xi_i(t)\xi_j(t') \rangle = \sigma^2 \delta_{ij}\delta(t-t')$$

(1)

where $\bar{r}_i$ is the position of particle $i$, $\sum_{j\in \text{VN}(i)}$ means summing over the nearest neighbors of particle $i$ determined by the Voronoi diagram [22], $\alpha$ is the strength of the orientational force, the strength of the attractive force $\varepsilon = 0.5$ if the particle is on the boundary and 0 otherwise, the reduced local average speed $u_i = \left|\sum_{j\in \text{VN}(i)} \bar{v}_j(t)\right|/(n_i v_0)$, where $n_i$ is the number of nearest neighbors of particle $i$, and the center-of-mass position of the nearest neighbors of particle $i$ is $\bar{r}_i^c = \frac{1}{n_i}\sum_{j\in \text{VN}(i)} \bar{r}_j$. The velocity of the particle at time $t$ is updated by $\bar{v}_i(t) = v_0 \hat{n}(\theta_i(t)) + 2v_0 \bar{F}_i^r(t)$, where $\hat{n}(\theta_i)$ is the normal vector of $\theta_i$, and the repulsive force

$$\bar{F}_i^r(t) = \sum_{j\in \text{VN}(i)} \hat{r}_{ij}(t-\Delta t) / \left(1+\exp\left(\frac{r_{ij}(t-\Delta t)}{0.5}-1\right)\right), \quad (2)$$

where $\hat{r}_{ij} = \bar{r}_{ij}/r_{ij}$, $r_{ij} = |\bar{r}_i - \bar{r}_j|$. Finally, the position of the particle is updated by $\bar{r}_i(t) = \bar{r}_i(t-\Delta t) + \bar{v}_i(t)\Delta t$ with $\Delta t = 0.05$.

To distinguish different phases, a *motion order parameter* (MOP) is defined as $\lambda = \varphi - |\omega|$, where $\varphi = |\langle \bar{v}_i(t) \rangle|/v_0$ is the reduced average speed and $\omega = \left\langle \frac{(\bar{r}_i(t) - \langle \bar{r}_i(t)\rangle) \times \bar{v}_i(t)}{|\bar{r}_i(t) - \langle \bar{r}_i(t)\rangle|v_0} \right\rangle$ is the reduced angular momentum. The range of $\varphi$ is [0, 1], $\omega$ is [-1, 1], and $\lambda$ is [-1, 1]. A state is identified as flocking if $\lambda > 1/3$, vortical if $\lambda < -1/3$, and disordered otherwise (see the Appendix).

A flock with 1000 particles has been simulated by the above model with $\alpha=0.02$ and $\sigma=1.12$ for $1.6\times 10^7$ steps, when either the flocking mode or the vortical mode can be stable for a period of time (~$10^5$ steps) before it spontaneously switches to the other mode. The time evolution of $\lambda$ during a simulation is plotted in Fig. 2. This simulation procedure was repeated with different random initial configurations for 24 times, and a sampling interval of 500 steps was used to collect simulation data.

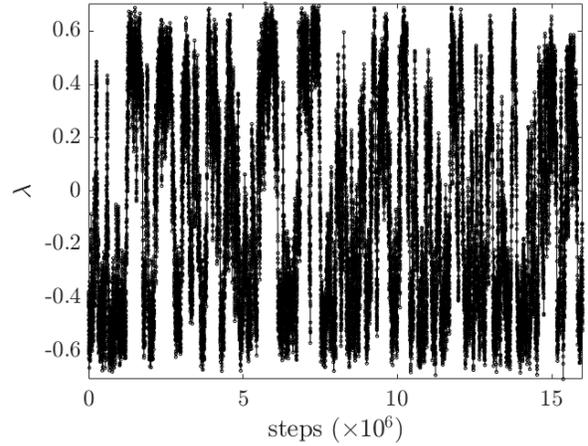

FIG. 2. Time evolution of $\lambda$ during one simulation.

As illustrated in Fig. 1, visual examination of the simulation trajectories reveals the following switching dynamics between the two modes. In the globally flocking state (Fig. 1(a)), some local vortices instantaneously appear and disappear due to thermal fluctuation induced by $\xi_i$ before a large enough vortex is generated and grows up quickly, turning the whole system into the globally vortical state (Fig. 1(b)). Reversely, starting from the vortical state (Fig. 1(b)), some local flocking pieces instantaneously appear and disappear before entering a metastable state with almost half of the particles forming a large flocking nucleus and the others still vortical (Fig. 1(c)), and after a while the system goes out of the metastable state to form the global flocking state (Fig. 1(a)).

The physical features and microscopic mechanisms of the above dynamical procedures are studied from the viewpoint of the CNT. A local cluster with two or more particles is identified as a flocking nucleus if the MOP of the cluster $\lambda_c > 1/3$, and as a vortical nucleus if $\lambda_c < -1/3$. To analyze the forward mode-switching dynamics, the size distribution of vortical nuclei $p_v(n)$ is calculated and plotted in Fig. 3(a). All simulation data are used in the calculation and the flocking motion is treated as "background" with respect to vortical nuclei. The corresponding free energy landscape $\Delta G_v \sim -\ln(p_v)$ is plotted in Fig. 3(b). There is only one maximum, indicating that it is a one-step nucleation process with only one free energy barrier to overcome. Under the framework of the CNT, we have found that the free energy landscape can be perfectly fitted by

$$\Delta G_v(n) = \gamma_{v,0} n^{g_{v,0}} - \gamma_{v,1} n^{-g_{v,1}} - \Delta \mu_v \exp(g_{v,2} n), \quad (3)$$

where $\gamma_{v,0} = 10.1$, $g_{v,0} = 0.0546$, $\gamma_{v,1} = 7.98$, $g_{v,1} = 0.357$, $\Delta \mu_v = 0.0119$, and $g_{v,2} = 0.00523$. The first two terms roughly correspond to the surface energy while the third term to the bulk energy. By locating the maximal value of the fitted curve, we identify that the critical nucleus size is 651.

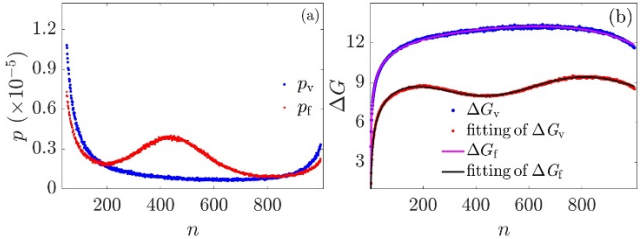

FIG. 3. (a) Size distributions of vortical and flocking nuclei. (b) Corresponding free energy landscapes along with their fitted curves.

The backward mode-switching dynamics from vortical to flocking is similarly analyzed by treating the vortical motion as the background with respect to the flocking nuclei. The size distribution and corresponding free energy landscape of flocking nuclei are also plotted in Figs. 3(a) and 3(b). The free energy landscape is fitted by

$$\Delta G_f(n) = \gamma_{f,0} n^{g_{f,0}} - \gamma_{f,1} n^{-g_{f,1}} - \Delta \mu_{f,1} \exp(g_{f,2} n) - \Delta \mu_{f,2} \exp\left(\frac{-(n-n_m)^2}{g_{f,3}}\right), \quad (3)$$

where $\gamma_{f,0} = 15.6$, $g_{f,0} = 0.0559$, $\gamma_{f,1} = 14.1$, $g_{f,1} = 0.0785$, $\Delta \mu_{f,1} = 1.11$, $g_{f,2} = 1.69 \times 10^{-3}$, $\Delta \mu_{f,2} = 2.89$, $g_{f,3} = 8.75 \times 10^4$, and $n_m = 440$ located at the local minimum of the free energy landscape. The first two terms roughly correspond to the surface energy, while the third term to the bulk energy. The fourth term reflects the appearance of the metastable state. The two local maxima of the free energy landscape locate respectively at 187 and 766, indicating that the backward mode switch is a two-step nucleation process.

By computer simulation with a Vicsek-like model, we have revealed that the dynamics of the forward and backward spontaneous switches between the flocking mode and the vortical mode of a finite-size active matter are asymmetric. The forward mode switch from flocking to vortical is a one-step nucleation process. Small vortical nuclei appear and disappear instantaneously in the flocking state due to thermal fluctuation until one larger than the critical nucleus grows up to turn the whole system into the vortex state by climbing over the single free energy barrier. Reversely, the backward mode switch from vortical to flocking is a two-step nucleation process. In the first stage, local pieces of small flocking nuclei generate in the vortical state until the system goes over the first free energy barrier to enter the metastable state with almost half of the particles becoming flocking. In the second stage, the system goes over the second barrier to become globally flocking. The physical reason causing the asymmetry of the mode-switching dynamics is that a flocking nucleus is easier to form than a vortical one due to the orientational force, so the forward procedure has to form a very large vortical nucleus simultaneously to fulfil the switch, while the backward one creates flocking structures in two sequential steps to have a lower overall switching free energy barrier in comparison with the case of one-step nucleation. This discovery deepens our understanding of motion dynamics in active matter and is anticipated to guide the efficient manipulations of artificial active matter, such as robots and UAVs.

This work was supported by the National Natural Science Foundation of China (No. 11947302). The allocation of computer time on the HPC cluster of ITP-CAS is also appreciated.

———

* Corresponding author: wangyt@itp.ac.cn.

## APPENDIX: Thresholds of the MOP

To determine the appropriate threshold of the MOP $\lambda$ for the flocking state, the system was simulated with $\alpha=0.1$ and various $\sigma$ values to observe the change of $\langle|\lambda|\rangle$ from the flocking state to the disordered state. As shown in Fig. 4(a), $\langle|\lambda|\rangle$ monotonically and continuously decreases with increasing $\sigma$, and its derivative reaches the minimal value at $\sigma = 2.50$, corresponding to $\langle|\lambda|\rangle=1/3$. Therefore, we regard the system as in the flocking state if $\lambda > 1/3$. Similarly, simulations from the vortical state to the disordered state with $\alpha=0.01$ and various $\sigma$ values were performed. As shown in Fig. 4(b), $\langle|\lambda|\rangle$ changes most rapidly at $\sigma = 0.882$ with corresponding $\langle|\lambda|\rangle=1/3$.

Consequently, the system is regarded as in the vortical phase if $\lambda < -1/3$.

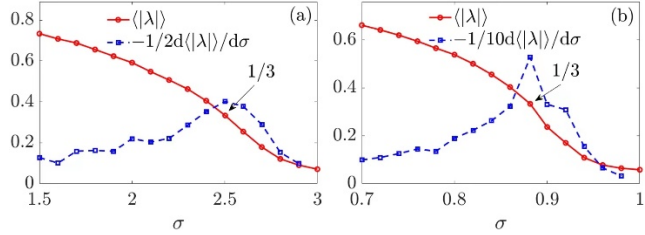

FIG. 4. $\langle |\lambda| \rangle$ and its derivative versus $\sigma$ with $\alpha$=0.1 from the flocking state to the disordered state (a), and with $\alpha$=0.01 from the vortical state to the disordered state (b).